\documentclass[a4paper,11pt]{article}
\usepackage{pos}
\usepackage{graphicx}
\usepackage{amsmath}

\title{
Manifestations of CP Violation in the B--Meson System: Theoretical Perspective}

\author*[a,b]{Eleftheria Malami}

\affiliation[a]{Center for Particle Physics Siegen (CPPS), Theoretische Physik 1, Universit\"at Siegen,\\
   D-57068 Siegen, Germany}

\affiliation[b]{Nikhef,\\
Science Park 105, 1098 XG Amsterdam, Netherlands}

\emailAdd{Eleftheria.Malami@uni-siegen.de}
\emailAdd{emalami@nikhef.nl}

\abstract{CP violation in the field of $B$ physics is a crucial topic for exploring the quark sector and search for New Physics, both for theorists and experimentalists. CP violation manifests itself in various ways and in this presentation, we will categorise the decays based on their different dynamics. We aim to present highlights related to the studies of CP violation in each category from a theoretical perspective.}

\FullConference{20th International Conference on B-Physics at Frontier Machines (Beauty2023)\\
 3-7 July, 2023\\
Clermont-Ferrand, France\\}

\begin{document}
\maketitle

\section{A Brief Introduction}
The $B$-meson system provides tests of the Standard Model (SM) and information for New Physics (NP) searches. A key point is that one can probe very high scales for NP, much higher than those in direct searches at colliders. We are dealing with precision physics, thus we perform indirect searches. Central role in the studies of $B$ physics plays the violation of the CP symmetry, which refers to the non-invariance of the weak interactions with respect to a combined charge-conjugation (C) and parity (P) transformation. Highlighting the key points of CP violation, we give an overview of the CP violation in the $B$--meson system. 

CP violation was discovered in 1964 through the observation of the $K_{L} \to \pi^+ \pi^-$ decay and is nowadays established in the kaon, $B$-meson and $D$-meson systems. As it comes in different manifestations, we categorise the $B$ decays based on their different dynamics, specifically according to the topologies that they originate from at leading order. In general, there are either tree or loop topologies, like penguins and boxes. Firstly, we discuss pure tree decays, such as $B_s \rightarrow D_s^{\pm} K^{\mp}$ and related modes. Then, we move on to transitions dominated by trees but also with penguin contributions. Key examples here are the $B_d \rightarrow J/\psi K_S$ and $B_s \rightarrow J/\psi \phi$ transitions. The third category refers to decays that are penguin dominated, such as the $B \rightarrow \pi K$ and the $B_s \rightarrow K^+ K^-$ systems. Lastly, there are decays which arise from electro-weak (EW) penguins and box topologies.

\section{Pure Tree Decays}
\begin{figure}[t!]
\centering
\includegraphics[width=0.26\linewidth]{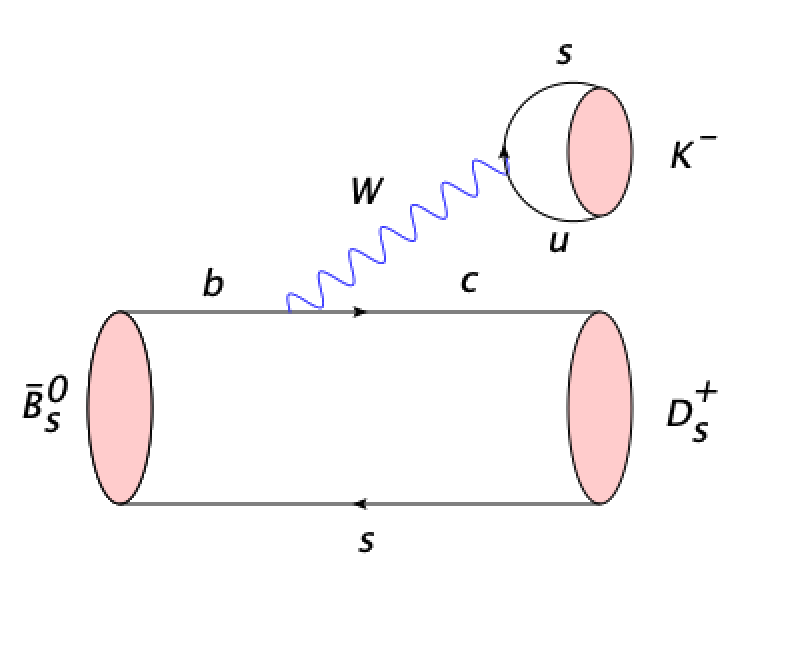}
\includegraphics[width=0.305\linewidth]{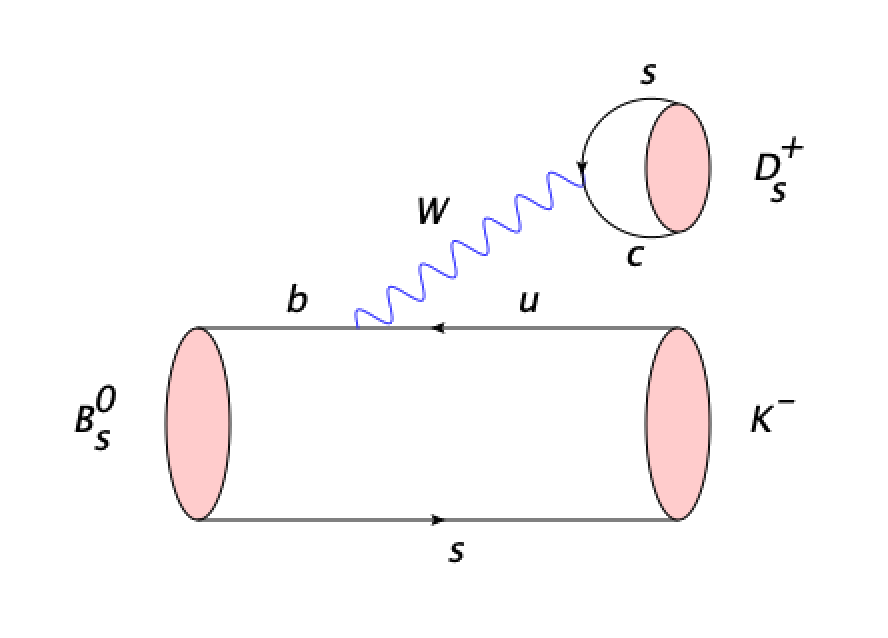} \\
\vspace{-0.4cm} 
\par\noindent\rule{\textwidth}{0.2pt}
\includegraphics[width=0.24\linewidth]{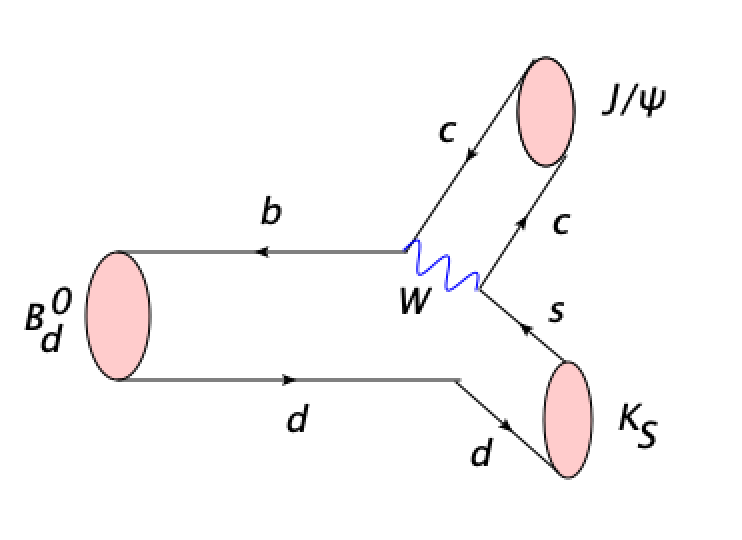} \hspace{-0.5cm}
\includegraphics[width=0.24\linewidth]{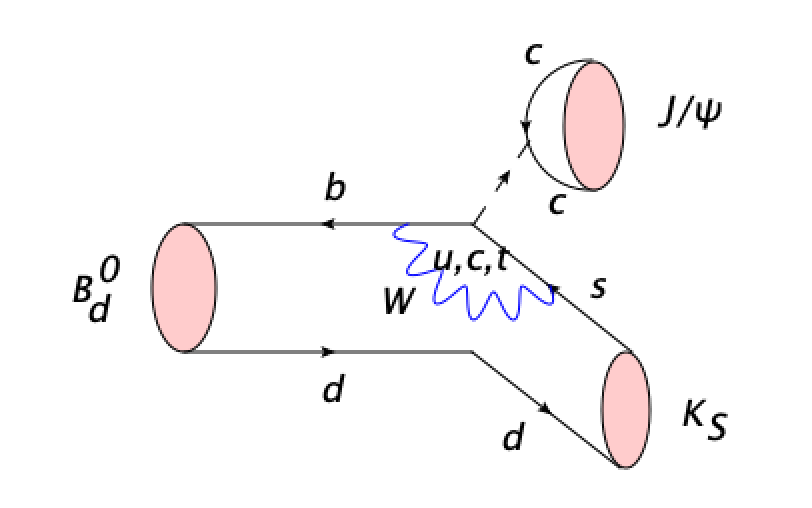} \vline  \hspace{0.4cm}
\includegraphics[width=0.233\linewidth]{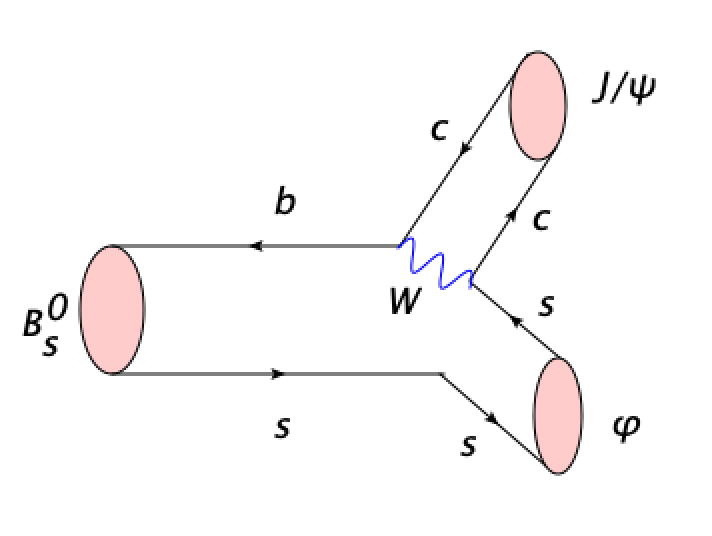} \hspace{-0.5cm}
\includegraphics[width=0.24\linewidth]{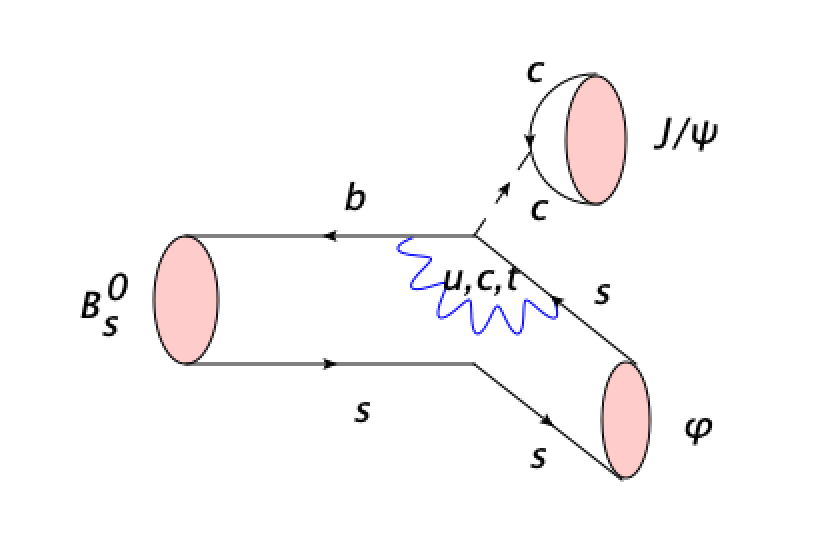} \\ 
\vspace{-0.4cm} 
\par\noindent\rule{\textwidth}{0.2pt}
\includegraphics[width=0.245\linewidth]{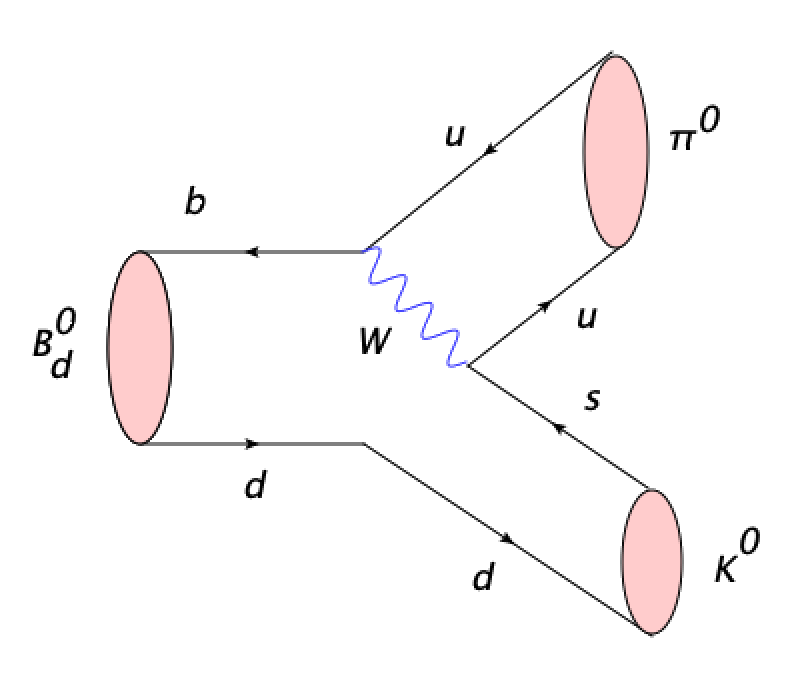}
\includegraphics[width=0.245\linewidth]{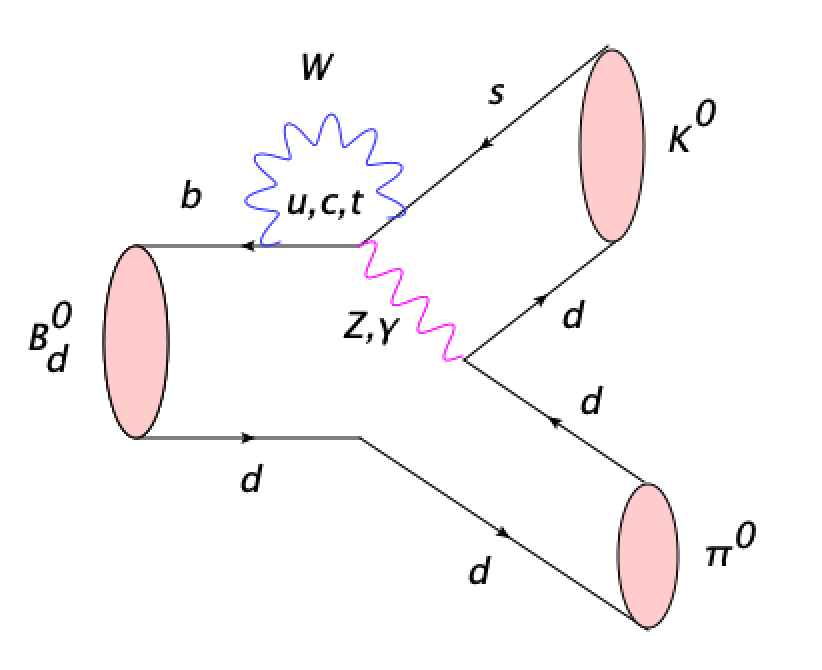}
\includegraphics[width=0.247\linewidth]{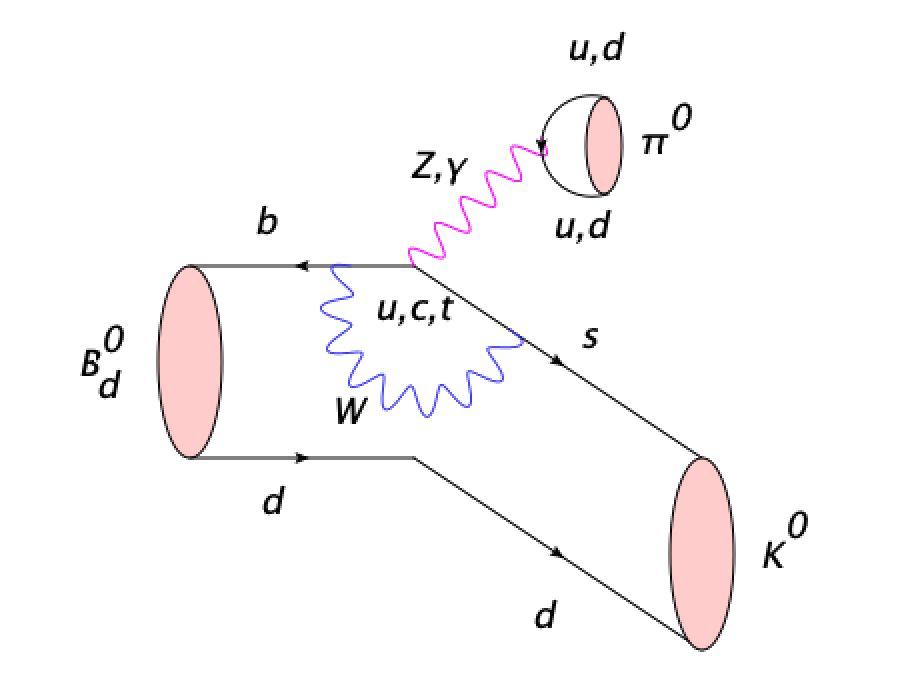}
\includegraphics[width=0.245\linewidth]{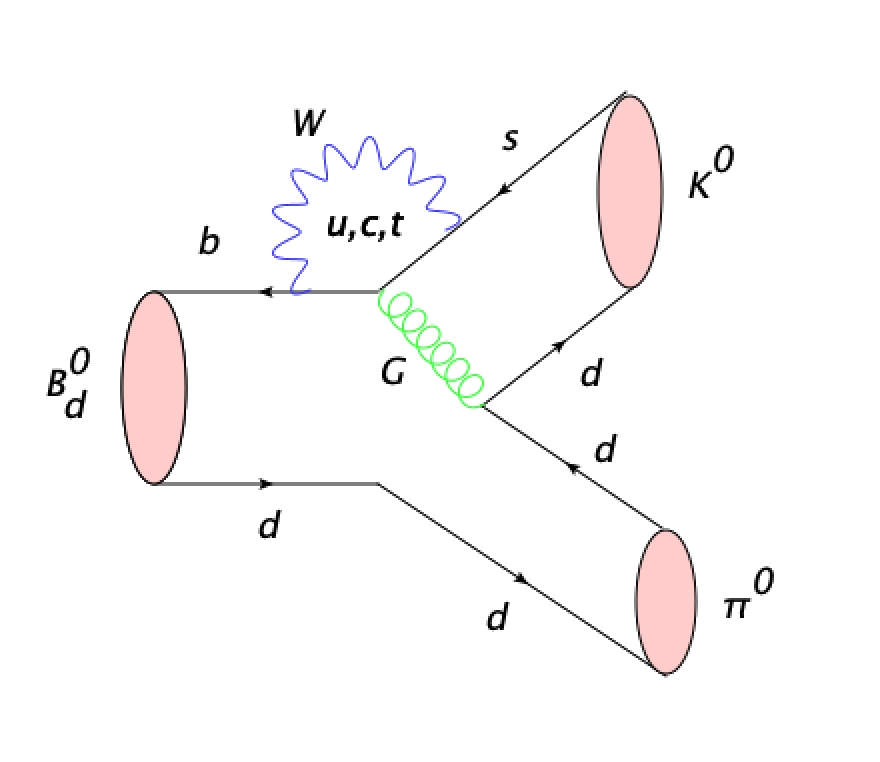} \\
\vspace{-0.4cm} 
\par\noindent\rule{\textwidth}{0.2pt}
\includegraphics[width=0.33\linewidth]{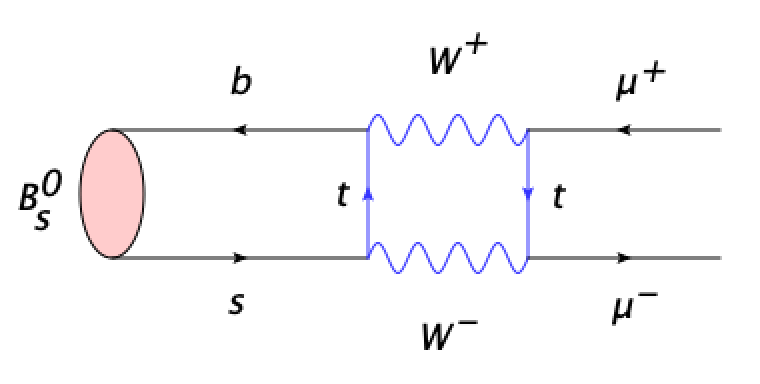} \hspace{0.6cm}
\includegraphics[width=0.33\linewidth]{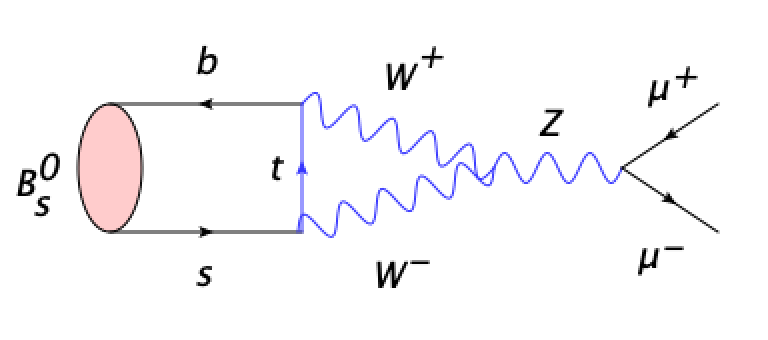}
\caption{Examples of topologies categorising the decays according to their different dynamics. Top row: Purely tree decays using as an example the $B_s \to D_s^{\pm} K^{\mp}$ system. Second row: Decays with tree and penguin topologies, such as the $B^0_d \to J/ \psi K^0_s$ and $B^0_s \to J/ \psi \phi$ mode. Third row: Penguin dominated decays like the $B^0_d$ $\rightarrow$ $\pi^0 K_s$ channel, showing firstly the tree diagram, then the colour--suppressed and the colour--allowed EW penguins as well as the QCD penguin. Bottom row: Box and penguin topologies for transitions like $B^0_s \to \mu^+ \mu^-$.}
\label{fig:tree}
\end{figure}

The first category of decays we present are those originating solely from tree topologies. More specifically, we focus on the $B_s \rightarrow D_s^{\pm} K^{\mp}$ decays and the associated puzzling anomalies. The corresponding Feynman diagrams are given in the top panel of Fig.\ref{fig:tree}. Due to $B^0_s -\bar{B}^0_s$ mixing, interference effects arise between the $\bar{B}^0_s\to D_s^+K^-$ and ${B}^0_s\to D_s^+K^-$ decays, leading to the following
time--dependent CP asymmetry \cite{RF-BsDsK}:
\begin{equation}
{\frac{\Gamma(B^0_s(t) \rightarrow f)-\Gamma(\bar{B}^0_s(t) \rightarrow {f})}{\Gamma(B^0_s(t) \rightarrow f)+\Gamma(\bar{B}^0_s(t) \rightarrow {f})}=\left[ \frac{{{C}} \ {{\cos(\Delta M_s \  t)}} + {{S}} \ {\sin(\Delta M_s \  t)}}{{{\cosh(\Delta \Gamma_s \  t/2)}} + {{\mathcal{A}_{\Delta \Gamma}}} {{\sinh(\Delta \Gamma_s \  t/2)}}} \right]}.
\label{eq:asym}
\end{equation}

The observables $C, \bar{C}, S, \bar{S}, {\cal A}_{\Delta \Gamma}, {\cal{\bar{A}}}_{\Delta \Gamma} $ allow a theoretically clean determination of the angle $\gamma$ of the Unitarity Triangle (UT). How do we determine this angle? We utilise the key relation:
\begin{equation}
\xi \times \bar{\xi}  = e^{-i2 (\phi_s + \gamma)},
\label{eq:gamma}
\end{equation}
where $\xi$ and $\bar{\xi}$ are observables that measure the strength of the correspodning interference effects. Following fro Eq.~\ref{eq:asym}, these quantities are determined as:
\begin{equation}
 C=({1-|\xi|^2})/({1+|\xi|^2}),  \quad S= ({2\,\text{Im}{\,\xi}})/({1 + |\xi|^2}), \quad 
 \mathcal{A}_{\Delta \Gamma}=({2\,\text{Re}\,\xi})/({1+|\xi|^2}).
\end{equation}
They are measured by the LHCb Collaboration, using the assumption that $C=-\bar{C}$, which holds in the SM. Here, we use the values from Ref.~\cite{BsDsK-LHCb-CP} and extract $\xi$ and $\bar{\xi}$. The mixing phase $\phi_s$ is determined through the $B_s \to J/\psi \phi$ modes and including penguin effects as in \cite{Barel:2020jvf}, the most updated value is $\phi_s=(-3.0\pm1.1)^\circ$. Consequently the only unknown value in Eq.~\ref{eq:gamma} is the $\gamma$ angle, which now can be determined. The values of Ref.~\cite{BsDsK-LHCb-CP} lead to a value of $\gamma$ which is much higher than the regime of $70^{\circ}$ \cite{Amhis:2019ckw}. This value is determined to be $\gamma=\left(131^{+17}_{-22}\right)^\circ $, suggesting a tension with the SM at the $3 \sigma$ level. Such an intriguing value shall be further explored and a detailed analysis can be found in Refs.~\cite{Fleischer:2021cwb,Fleischer:2021cct}. Could this suggest NP? 

If there are NP effects entering at the amplitude level, they should also manifest themselves in the branching ratios. Thus, we extract ``theoretical'' individual branching ratios, which refer to the branching fractions at time $t=0$, where mixing effects are ``switched off'':
\begin{align}\label{BRbar-Ds+K-}
\mathcal{B}(\bar{B}^0_s \to D_s^+K^-)_{\text{th}} &=2 \left[{|\xi|^2}/{\left(1+|\xi|^2 \right)} \right]\mathcal{B}_{\text{th}} = (1.94 \pm 0.21) \times 10^{-4}, \\
\mathcal{B}(B^0_s\to D_s^+K^-)_{\text{th}} &=2 \left[{1}/{\left(1+|\xi|^2 \right)} \right]\mathcal{B}_{\text{th}} =(0.26 \pm 0.12) \times 10^{-4}, \\
{\text{where  \ }} 
 \mathcal{B}_{\text{th}} =  \bar{\mathcal{B}}_{\text{th}} &=
\left[\frac{1-y_s^2}{1+y_s\langle {\cal A}_{\Delta\Gamma}\rangle_+}\right]\langle\mathcal{B}_{\text{exp}}\rangle = (1.10 \pm 0.09) \times 10^{-4} 
\end{align}
\begin{equation}
{\text{with  \ }} \langle\mathcal{B}_{\text{exp}}\rangle \equiv \frac{1}{2}\left(\mathcal{B}_{\text{exp}} + \bar{\mathcal{B}}_{\text{exp}}\right)=
\frac{1}{2} \, \mathcal{B}^{\text{exp}}_\Sigma =\frac{1}{2} \ (2.27 \pm 0.19) \times 10^{-4}.
\end{equation}
The $\mathcal{B}^{\text{exp}}_\Sigma$ experimental value is given in Ref.~\cite{ParticleDataGroup:2022pth}.

Key quantity is the phenomenological colour factor $|a_1|$. In order to determine the theoretical values of  $|a_1|$, we utilise the framework of factorisation, which is expected to work very well for the $b \to c$ modes and the channel $B_s \to D_s^+ K^-$ is a prime example.  We write the factorised amplitude in terms of CKM matrix elements, the kaon decay constant, the corresponding hadronic form factor and the parameter
\begin{equation}\label{a-eff-1-DsK}
a_{\rm 1 \, eff }^{D_s K}=a_{1}^{D_s K} \left(1+{E_{D_s K}}/{T_{D_s K}}\right),
\end{equation}
which takes the exchange topologies into account. More specifically, the $a_{1}^{D_s K}$ factor refers to non-factorisable effects entering the tree topologies, the $T_{D_s K}$ stands for these colour-allowed tree amplitudes while  the $E_{D_s K}$ describes the non-factorisable exchange topologies. The current state-of-the-art values are $|a_1| \approx 1.07$ with uncertainties at the percent level. Exchange topologies for the $B_s \to D_s K$ system do not suggest any anomalous enhancement. For the calculation of the experimental values of $|a_1|$, the cleanest theoretical method is to create ratios of branching fractions with the semileptonic partner decays, such as:
\begin{equation}
 \frac{\mathcal{B}(\bar{B}^0_s \rightarrow D_s^{+}K^{-})_{\rm th}}{{\mathrm{d}\mathcal{B}\left(\bar{B}^0_s \rightarrow D_s^{+}\ell^{-} \bar{\nu}_{\ell} \right)/{\mathrm{d}q^2}}|_{q^2=m_{K}^2}}=6 \pi^2 f_{K}^2 |V_{us}|^2 |a_{\rm 1 \, eff }^{D_s K}|^2  {\Phi_{\text{ph}}} \left[ \frac{{F_0^{B_s \rightarrow D_s}(m_K^2)}}{{F_1^{B_s \rightarrow D_s}(m_K^2)}} \right]^2, \  \ {\Phi_{\text{ph}}} \approx 1,
    \label{semi}
\end{equation}
which includes the matrix element  $|V_{us}|$, the kaon decay constant $f_{K}$ and ratios of form factors. Analogous ratios can be written for other decays with similar dynamics. Following these lines, we obtain the experimental  $|a_1|$ results for the $b \to c$ modes:
\begin{align}
\bar{B}^0_s \rightarrow D_s^+K^- {\text{decay:}} \quad &|a_{\rm 1}^{D_d K}|=0.82\pm0.11, \ \  \ \bar{B}^0_d \rightarrow D_d^+K^- {\text{decay:}} \quad |a_{\rm 1}^{D_d K}|=0.83\pm0.05, \nonumber \\ 
\bar{B}^0_d\to D_d^+\pi^-  {\text{decay:}} \quad  &|a_1^{D_d \pi}|=0.83\pm 0.07, \quad \bar{B}^0_s\to D_s^+\pi^-  {\text{decay:}} \quad  |a_1^{D_s\pi}|=0.87\pm0.06.  \nonumber 
\end{align} 
In an analogous way, we work for the $b \to u$ modes and obtain
\begin{align}
\bar{B}^0_s\to K^+ D_s^-  {\text{decay:}} \quad  &|a_{\rm 1}^{K D_s}| =0.77 \pm 0.19, \quad \bar{B}^0_d\to \pi^+D_s^-  {\text{decay:}} \quad  |a_{\rm 1}^{\pi D_s}| = 0.78\pm0.05. \nonumber
\end{align} 
Comparing these results with the theoretical predictions, we observe that they are all much smaller than the theoretical values, showing tensions up to the $4.8$~$\sigma$ level. A pattern which holds also for the $b \to u$ channels, where in principle, factorisation is on less solid ground. This situation with the $|a_1|$ values marks the second intriguing case in the $B_s \to D_s^{\pm} K^{\mp}$ system. These two puzzles complement each other.  The possibility of NP effects in non-leptonic decays is exciting and discussions can be found in Refs.~\cite{Lenz:2019lvd,Iguro:2020ndk, Cai:2021mlt}.

Therefore, the next step is to move towards studies of New Physics (NP). For this purpose, we introduce the NP parameters:
\begin{equation}
\bar{\rho} \, e^{i \bar{\delta}} e^{i \bar{\varphi}}  \equiv { A(\bar{B}^0_s \rightarrow D_s^+ K^-)_{{\text{NP}}} }/{  A(\bar{B}^0_s \rightarrow D_s^+ K^-)_{{\text{SM}}} }, 
\end{equation}
with $\bar{\delta}$ and $\bar{\varphi}$ denoting the CP-conserving and CP-violating phases, respectively (and ${\rho},{\delta},{\varphi}$ for the CP-conjugate case). Now we generalize Eq.~\ref{eq:gamma} and obtain:
\begin{equation}
\xi \times \bar{\xi}  = \sqrt{1-2\left[\frac{C+\bar{C}}{\left(1+C\right)\left(1+\bar{C}\right)}
\right]}e^{-i\left[2 (\phi_s +\gamma_{\rm eff})\right]},
\label{eq:generxi}
\end{equation}
which is theoretical clean and includes an an ``effective'' angle 
\begin{equation}
\gamma_{\rm eff}\equiv \gamma + \gamma_{\text{NP}} = (131^{+17}_{-22})^\circ, \quad {\text{where\ }} \gamma_{\text{NP}}=f(\rho, \bar{\rho}, \varphi, \bar{\varphi}) . 
\end{equation}
Setting  $\delta=\bar{\delta}=0$, correlations between the NP parameters can be determined through the following formulas: 
\begin{equation}
  \tan{\Delta \phi} = \frac{\rho \sin{\phi} + \bar{\rho} \sin{\bar{\phi}} + \rho \bar{\rho} \sin{(\bar{\phi} + \phi )}}{1 + \rho \cos{\phi} + \bar{\rho} \cos{\bar{\phi}} + \rho \bar{\rho} \cos{(\bar{\phi} + \phi)}}, {\text{\ \ where\ }} \Delta \phi= -(61\pm20)^\circ,
  \label{eq:tan}
\end{equation}
\begin{align}
    b&= 1 + 2 \rho \cos{\delta} \cos{\phi} + \rho^2 = \frac{{\mathcal{B}(\bar{B}^0_s \rightarrow D_s^{+}K^{-})_{\rm th}}/ \left[{{\mathrm{d}\mathcal{B}\left(\bar{B}^0_s \rightarrow D_s^{+}\ell^{-} \bar{\nu}_{\ell} \right)/{\mathrm{d}q^2}}|_{q^2=m_{K}^2}} \right]}
    {6 \pi^2 f_{K}^2 |V_{us}|^2 |a_{1}^{D_s K}|^2 X_{D_s K}},
     \label{betabar}
\end{align}
where $|a_{1}^{D_s K}|$ is now an input parameter. An analogous relation can be written for $\bar{b}$ and both quantities can be determined: 
\begin{equation}
b=0.58 \pm 0.16, \qquad \bar{b} = 0.50 \pm 0.26,
\end{equation}
showing tension with the SM value of 1. This model--independent strategy, which is discussed in detail in Refs.~\cite{Fleischer:2021cwb,Fleischer:2021cct}, indicates that the data could be accommodated with NP contributions at the $30 \%$ level. It is worth noting that a new standalone LHCb Run II measurement has recently been reported \cite{LHCb:2023mcw} and it should also be further explored. This strategy may lead to the establishment of new sources of CP violation in the future.

\section{Decays Dominated by Trees But Also Involving Penguin Contributions}
\label{Sec:two}
The $B^0_d \to J/ \psi K^0_s$ and $B^0_s \to J/ \psi \phi$ decays have been characterised as the golden modes for establishing CP violation in the $B$ system and have historically received a lot of attention. Nowadays, with impressive experimental progress, we have reached the level of precision where it is important to start including the penguin contributions.

The Feynman diagrams describing these processes are depicted in the second row of Fig.\ref{fig:tree}, illustrating the contributions of the colour suppressed tree diagrams as well as the penguin topologies, which are doubly-Cabibbo suppressed. This entails that the decay amplitude is proportional to the term $\lambda^2$, where $\lambda \equiv |V_{us}| \approx 0.22$ is the Wolfenstein parameter. Consequently, calculating the penguins is very difficult.

The central role for the analysis here is played by the mixing phases $\phi_s$ and $\phi_d$. Due to contributions from the doubly Cabibbo-suppressed penguins, a hadronic phase shift $\Delta \phi_q$ is introduced and we measure an effective 
phase $\phi_q^{\text{eff}}$ defined as follows:
\begin{equation}
\phi_q^{\text{eff}}= \phi_q + \Delta \phi_q= \phi_q^{\text{SM}} + \phi_q^{\text{NP}} + \Delta \phi_q.
\end{equation}
Here $\phi_q$ is the phase that we access experimentally and consists of the SM part, which is determined through the Unitarity Triangle (UT) and the NP part which includes effects from CP violation from the beyond the SM. The parameter $\Delta \phi_q$ depends on penguin parameters, the $a$ and $\theta$, and provides a measure of the ratio of penguin over tree contributions.

Since the hadronic effects that characterise the non-leptonic $B^0_d \to J/ \psi K^0_s$ and $B^0_s \to J/ \psi \phi$ systems are difficult to calculate in QCD, as non-perturbative, we follow a different strategy, as presented in Ref.~\cite{Barel:2020jvf}. We utilise control channels, where the hadronic effects are not doubly Cabibbo--suppressed. More specifically, applying the SU(3) flavour symmetry of strong interaction, the penguin parameters of the $\bar{b} \to \bar{s} c \bar{c}$  and $\bar{b} \to \bar{d} c \bar{c}$ transitions are related as follows:
\begin{equation}\label{eq:su3_relation}
a' e^{i \theta'}= a e^{i \theta}\:.
\end{equation}
The partner control channels of the $B^0_d \to J/\psi K^0_S$ decay are the $B^0_s \to J/\psi K^0_S$ and  $B^0_d \to J/\psi \pi^0$ channels while for $B^0_s \to J/\psi \phi$ we have the $B^0_d \to J/\psi \rho^0$ channel. Therefore, we utilise all these five channels and we make use of their CP asymmetries, using the relation:
\begin{equation}
  \sin\left(\phi_q^{\text{eff}}\right)
     = {\eta_f \mathcal{A}_{\text{CP}}^{\text{mix}}(B_q\to f)} / {\sqrt{1 - \left(\mathcal{A}_{\text{CP}}^{\text{dir}}(B_q\to f)\right)^2}} \:,
\end{equation}
where $\eta_f$ is the CP eigenvalue of the final state $f$. A simultaneous fit for the penguin parameters and mixing phases from the CP asymmetries of all these $B_s \to J/\psi X$ channels, where we properly take into account the dependencies between $\phi_d$, $\Delta \phi_d$, $\phi_s$ and $\Delta \phi_s$, leads to the extraction of the corresponding hadronic parameters and the mixing phases:
\begin{equation}\label{eq:results_phiq}
  \phi_d = \left(45.4_{-1.1}^{+1.3}\right)^{\circ}  \:, \qquad
     \phi_s = \left(-3.0 \pm 1.1\right)^{\circ}\:.
\end{equation}
These are the most updated values, which take the penguin effects into account. Consequently, the main highlight is that this strategy includes the impact of the penguins on the CP asymmetries.

The second important point in the analysis of these decays is the clean determination of the colou--suppression factor $|a_2|$ with the help of ratios of branching fractions with partner semileptonic decays, in a similar way as it was already used in the case of the $B_s \to D_s^{\pm} K^{\mp} $ system and the $|a_1|$ determination. Thus, one can write \cite{Barel:2020jvf}:
\begin{equation}\label{eq:SL_ratio}
\frac{{\mathcal{B}(B^0_d \to J/\psi \pi^0)}}{{d\mathcal{B}/dq^2|_{q^2=m_{J/\psi}^2}(B^0_d \to \pi^- \ell^+ \nu_{\ell}) }} \propto  \  (1 - 2 a\cos\theta\cos\gamma + a^2) \times \left[ a_2 (B_d^0\to J/\psi\pi^0) \right]^2 \:,
\end{equation}
which allows the extraction of $|a_2|$. For this case, the obtained value is $|a_2|=0.363^{+0.066}_{-0.079}$, which agrees well with naive factorisation, suggesting that factorisation seems to work better than it was expected in this category of decays.

Last but not least, discussing important aspects of the $B^0_q$--$\bar B^0_q$ mixing phenomenon, we would like to emphasize interesting applications of $\phi_d$ and $\phi_s$ as well as highlight the topic of the UT apex determination. One way of determining the UT apex is utilising the angle $\gamma$ and the UT side $R_b$. However, in the extraction of $R_b$ tensions arise between various theoretical and experimental approaches. More specifically, there are discrepancies between the inclusive and exclusive determinations of the $|V_{ub}|$ and $|V_{cb}|$ matrix elements. Therefore, we advocate that it is important to perform the analysis separately for the inclusive and exclusive case and avoid making averages, in order to determine $R_b$ and consequently extract the UT apex.  In addition, a third possibility is studied in the literature, which is hybrid combination of exclusive $|V_{ub}|$ and inclusive $|V_{cb}|$. Studying the apex extraction for every case and on top of that, utilising the hyperbola following from $\varepsilon_K$, we come to the conclusion that the hybrid scenario is the one that provides the most consistent picture with the SM \cite{DeBruyn:2022zhw}. Therefore, in the future, it can be used to resolve the inclusive and exclusive puzzle. A key question is how big the space for NP in $B^0_q$--$\bar B^0_q$ mixing is. Utilising the mixing phases, NP contributions can be explored and parametrised,  performing fits for the parameters $\kappa_q$ and $\sigma_q$. These parameters denote the size of NP effects and the phase for additional CP violating effects, respectively. The results of these fits have interesting applications in rare leptonic decays, a category that we will discuss later. 

\section{Decays Dominated by Penguins}
Let us know discuss the $B \to \pi K $ and $B_{(s)} \to KK$ systems, where the main contributions come from penguin topologies. Firstly, regarding the $B \to \pi K $ decays, the most interesting channel for CP violation studies is $B_d^0 \to \pi^0 K_S$, as this is the only mode that exhibits mixing-induced CP violation. As illustrated in the third row of Fig.\ref{fig:tree}, the decay is dominated by QCD diagrams, but 
EW penguins play also an important role. Therefore, it is particularly interesting to measure CP violation in this channel with highest precision. 

\begin{figure}[t!]
\centering
\includegraphics[width=0.35\linewidth]{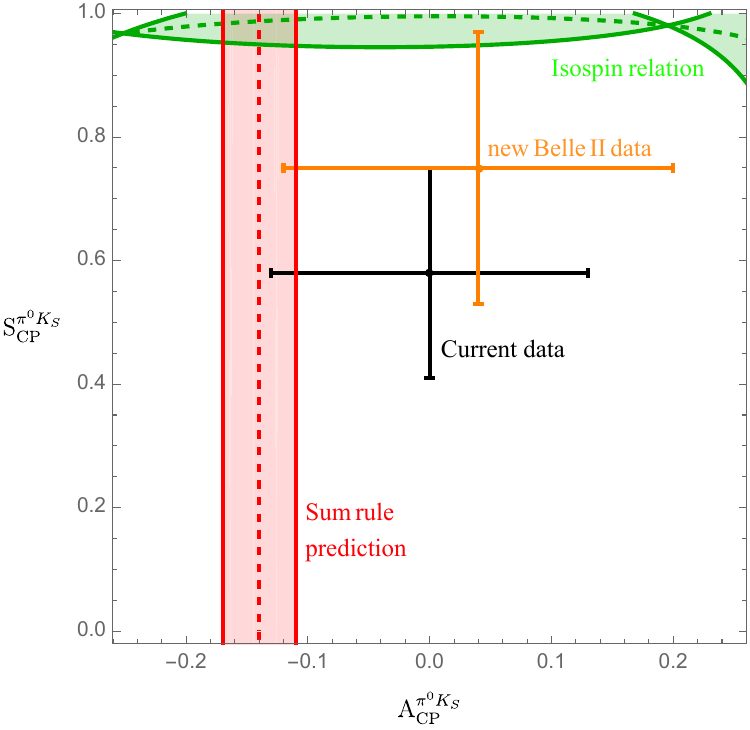}
\caption{Illustration of the puzzle related to CP asymmetries of the $B_d^0 \to \pi^0 K_S$ decay.}
\label{fig:two}
\end{figure}

Following the analysis presented in Refs.~\cite{Fleischer:2018bld,Fleischer:2008wb,Buras:2004ub}, isospin relations between decay amplitudes can be utilised in order to obtain correlations between the mixing-induced and the direct asymmetry of the $B_d^0 \to \pi^0 K_S$ channel. This correlation between the CP asymmetries is given in  Fig.~\ref{fig:two}. Comparing the contour that comes isospin analysis (green contour) with the current data for the CP asymmetries (black cross), we find tensions between them. In order for this puzzle to be resolved there either should be a change of data or NP effects in the EW penguin sector should be present. A new Belle II measurement \cite{Veronesi:2023dak} suggests a value of the CP asymmetries of the $B_d^0 \to \pi^0 K_S$ mode which seems to agree better with the isospin results within uncertainties (orange cross). This is an interesting point that should be further explored. 

Concerning the $B_s^0 \to K^-K^+$ decay, we have the first observation of CP violation by the LHCb Collaboration \cite{LHCb:2020byh}. The interesting finding is that there is a surprising difference between the direct CP asymmetries of the $B_s^0 \to K^-\pi^+$ and $B_d^0 \to \pi^-K^+$ decays. As discussed in Ref.~\cite{Fleischer:2022rkm}, exchange and penguin-annihilation topologies can accommodate this difference at the level of $20\%$. In the same paper, the determination of the $\gamma$ angle of UT is discussed using only CP asymmetries and no branching ratio information. The result of $\gamma=\left( 65^{+11}_{-7}\right)^\circ$ agrees excellently with the value coming from $B \to D K$ decays, which provides a clean
determination. Additionally, the phase $\phi_s$ can also be obtained through a new novel method using semileptonic $B^0_s$ and $B^0_d$ differential rates, highlighting again how useful the methodology with ratios with semileptonic partner decays is.

\section{Decays Arising from Electroweak Penguins and Box Topologies}
The last category refers to decays that arise from EW penguin and box topologies, such as the $B^0_d \to \mu^+  \mu^-$ and $B \to K \ell^+ \ell^-$ transitions, which have much simpler dynamics with respect to strong interactions than the non-leptonic decays. The corresponding topologies for the $B^0_s \to \mu^+ \mu^-$  decay are shown at the bottom row of Fig.~\ref{fig:tree}. 

Let us first refer back to the phenomenon of  $B^0_q$--$\bar B^0_q$ mixing that we discussed in Sec.~\ref{Sec:two} and provide more details regarding the applications of mixing on leptonic decays. NP studies depend strongly on both the UT apex and the $|V_{cb}|$ matrix element. Thus, in the determination of NP in the $B^0_s \to \mu^+ \mu^-$  decay, it is essential to manage to minimise the impact of the CKM parameters \cite{DeBruyn:2022zhw}. How can we do that? We create the ratio between the branching fraction of this decay and the mass difference $\Delta m_s$ \cite{Buras:2003td,Buras:2021nns}:
\begin{equation}
R_{s\mu}= \bar{\mathcal{{B}}}(B^0_s \to \mu^+ \mu^-)/ \Delta m_s,
\end{equation}
where the CKM elements drop out in the SM. We note though that we have to  take possible NP contributions in the $B^0_q$--$\bar B^0_q$ mixing into account, following the NP analysis in Ref.~\cite{DeBruyn:2022zhw}.  This allows us to constrain the pseudo-scalar and scalar parameters, $|P^s_{\mu \mu}|$ and $|S^s_{\mu \mu}|$ from the branching ratio and the ratio $R_{s\mu}$, thus it is an important point in the NP studies.

These rare decays have interesting phenomenology related to CP violation and the time dependent CP asymmetries have similar structure as those in the non-leptonic decays. More specifically, for the $B^0_{s,d} \to \ell^+ \ell^-$ decays, the CP--violating asymmetries coming from interference effects would be very interesting to measure but this is experimentally very challenging. Similarly, for $B^0_d \to K_S \mu^+ \mu^-$ channel, the interference effects through $B^0_d$--$\bar B^0_d$ mixing lead to mixing-induced CP violation. Usually for the CP--violating effects in  the NP analysis of rare decays, only real coefficients are considered. However, Wilson coefficients could also be complex. This case is really interesting to explore and a discussion can be found in Ref.~\cite{Fleischer:2022klb}.

An experimental highlight here is the December 2022 $R_K^{(*)}$ measurement \cite{LHCb:2022qnv,LHCb:2022vje}, which is now compatible with the SM predictions. In addition, the measured decay rates of the $B \to K \mu^+ \mu^-$ decays are too small and below the SM predictions at the $3.5~\sigma $ level. What could these imply for the Lepton Flavour Universality (LFU) violation? An analysis discussed in Refs.~\cite{Fleischer:2022klb,Fleischer:2023zeo} explores CP--violating effects in the NP analysis of rare decays considering complex Wilson coefficients. The proposed strategy suggests that starting from a complex muonic Wilson coefficient and using as input the new $\langle R_K \rangle$ measurement allows the determination of the electronic Wilson coefficients and consequently the extraction of the direct and mixing-induced CP asymmetry for the electronic modes \cite{Fleischer:2023zeo}. As a result, constraining the electronic NP Wilson coefficients, their magnitude and the mixing phases differ significantly from their muonic counterparts and the similar pattern follows for the resulting CP asymmetries between the electronic and the muonic channels. The conclusion is that despite the fact that the $R_K$ anomaly is now gone, if there is CP--violating NP entering, there is still significant violation of the electron-muon universality at the level of the Wilson coefficients. This is a key finding for searches of CP violation in the final state decays involving electrons and muons and consequently tests of LFU at the high-precision era.


\section{Outro}
CP violation continues to be a prime player for
exploring the flavour sector and New Physics searches 
for both theorists and experimentalists. Exciting times are ahead! 

\section*{Acknowledgements}
\noindent I would like to thank the organisers of Beauty 2023  for the invitation to participate in such an outstanding conference and the opportunity of great interactions and fruitful discussions with the fellow participants.


\begin{thebibliography}{99}
\bibitem{RF-BsDsK}
R.~Fleischer,
Nucl. Phys. B \textbf{671} (2003), 459-482

\bibitem{BsDsK-LHCb-CP}
R.~Aaij {\it et al.} [LHCb Collaboration],
  JHEP {\bf 1803}, 059 (2018)
  
\bibitem{Barel:2020jvf}
M.~Z.~Barel, K.~De Bruyn, R.~Fleischer and E.~Malami,
J. Phys. G \textbf{48} (2021) no.6, 065002

\bibitem{Amhis:2019ckw}
Y.~S.~Amhis \textit{et al.} [HFLAV],
Eur. Phys. J. C \textbf{81}, no.3, 226 (2021)

  
\bibitem{Fleischer:2021cwb}
R.~Fleischer and E.~Malami,
Eur. Phys. J. C \textbf{83} (2023) no.5, 420

\bibitem{Fleischer:2021cct}
R.~Fleischer and E.~Malami,
Phys. Rev. D \textbf{106} (2022) no.5, 056004

\bibitem{ParticleDataGroup:2022pth}
R.~L.~Workman \textit{et al.} [Particle Data Group],
PTEP \textbf{2022} (2022), 083C01

\bibitem{Lenz:2019lvd}
A.~Lenz and G.~Tetlalmatzi-Xolocotzi,
JHEP \textbf{07} (2020), 177

\bibitem{Iguro:2020ndk}
S.~Iguro and T.~Kitahara,
Phys. Rev. D \textbf{102} (2020) no.7, 071701

\bibitem{Cai:2021mlt}
F.~M.~Cai, W.~J.~Deng, X.~Q.~Li and Y.~D.~Yang,
JHEP \textbf{10} (2021), 235


\bibitem{LHCb:2023mcw}
 [LHCb],
LHCb-CONF-2023-004.

\bibitem{DeBruyn:2022zhw}
K.~De Bruyn, R.~Fleischer, E.~Malami and P.~van Vliet,
J. Phys. G \textbf{50} (2023) no.4, 045003

\bibitem{Fleischer:2018bld}
R.~Fleischer, R.~Jaarsma, E.~Malami and K.~K.~Vos,
Eur. Phys. J. C \textbf{78} (2018) no.11, 943

\bibitem{Fleischer:2008wb}
R.~Fleischer, S.~Jager, D.~Pirjol and J.~Zupan,
Phys. Rev. D \textbf{78} (2008), 111501

\bibitem{Buras:2004ub}
A.~J.~Buras, R.~Fleischer, S.~Recksiegel and F.~Schwab,
Nucl. Phys. B \textbf{697} (2004), 133-206

\bibitem{Veronesi:2023dak}
M.~Veronesi [BELLE II],
[arXiv:2305.09153 [hep-ex]].

\bibitem{LHCb:2020byh}
R.~Aaij \textit{et al.} [LHCb],
JHEP \textbf{03} (2021), 075


\bibitem{Fleischer:2022rkm}
R.~Fleischer, R.~Jaarsma and K.~K.~Vos,
JHEP \textbf{02} (2023), 081

\bibitem{Buras:2003td}
A.~J.~Buras,
Phys. Lett. B \textbf{566} (2003), 115-119

\bibitem{Buras:2021nns}
A.~J.~Buras and E.~Venturini,
Acta Phys. Polon. B \textbf{53} no.6, 6-A1

\bibitem{Fleischer:2022klb}
R.~Fleischer, E.~Malami, A.~Rehult and K.~K.~Vos,
JHEP \textbf{03} (2023), 113

\bibitem{LHCb:2022vje}
R.~Aaij \textit{et al.} [LHCb],
Phys. Rev. D \textbf{108} (2023) no.3, 032002

\bibitem{LHCb:2022qnv}
R.~Aaij \textit{et al.} [LHCb],
Phys. Rev. Lett. \textbf{131} (2023) no.5, 051803



\bibitem{Fleischer:2023zeo}
R.~Fleischer, E.~Malami, A.~Rehult and K.~K.~Vos,
JHEP \textbf{06} (2023), 033



\end{thebibliography}
\end{document}